\documentclass[british,longauth,structabstract]{aa}
\usepackage{mathptmx}
\usepackage[T1]{fontenc}
\usepackage{graphicx}
\usepackage{amssymb}
\usepackage[authoryear]{natbib}

\makeatletter

\def\lesssim{\mathrel{\hbox{\rlap{\hbox{\lower4pt\hbox{$\sim$}}}\hbox{$<$}}}}

\makeatother

\usepackage{babel}

\begin{document}

\title{Polarisation Observations of VY Canis Majoris H$_{2}$O $5_{32}-4_{41}$
620.701 GHz Maser Emission with HIFI\thanks{ Herschel is an ESA space observatory with science instruments provided by European-led Principal Investigator consortia and with important participation from NASA.}}


\author{Martin Harwit \inst{1}\and Martin Houde \inst{2}\and Paule Sonnentrucker
\inst{3}\and A. C. A. Boogert \inst{4}\and J. Cernicharo \inst{5}\and E. de Beck \inst{6}\and L. Decin \inst{6,16}\and C. Henkel \inst{7}\and R. D. Higgins \inst{8}\and W. Jellema \inst{9}\and
A. Kraus \inst{7}\and Carolyn M$^{\mathrm{c}}$Coey \inst{10,2}\and
G. J. Melnick \inst{11}\and K. M. Menten \inst{7}\and C. Risacher
\inst{9}\and D. Teyssier \inst{12}\and J. E. Vaillancourt \inst{13}\and
J. Alcolea \inst{14}\and V. Bujarrabal \inst{15}\and C. Dominik
\inst{16,17}\and K. Justtanont \inst{18}\and A. de Koter \inst{16,19}\and
A. P. Marston\inst{12}\and H. Olofsson \inst{18,20}\and P. Planesas
\inst{15,21}\and M. Schmidt \inst{22}\and F. L. Sch\"oier \inst{18}\and
R. Szczerba \inst{22}\and L. B. F. M. Waters \inst{6,16}}

\institute{Cornell University, Center for Radiophysics \& Space Research, 511
H street, SW, Washington, DC 20024-2725 USA \\ \email{harwit@verizon.net}\and
University of Western Ontario, Department of Physics and Astronomy,
London, Ontario, Canada N6A 3K7\and Johns Hopkins University, Department
of Physics and Astronomy, Baltimore, MD, 21218 USA\and IPAC, Caltech, Pasadena, CA 91925 USA\and Consejo Superior de Investigaciones Cientificas, 28006, Madrid, Spain\and Katolieke
Universiteit Leuven, Institut voor Sterrenkunde, Heverlee 3001 Belgium\and
Max-Planck-Institut f\"ur Radioastronomie, Auf dem H\"ugel 69, 53121 Bonn,
Germany\and National University of Ireland, Maynooth, Department
of Experimental Physics, County Kildare, Ireland\and Space Research
Organization of the Netherlands (SRON) 9700 AV Groningen, Netherlands\and
University of Waterloo, Department of Physics and Astronomy, Waterloo,
Ontario, Canada N2L 3G1\and Harvard-Smithsonian Center for Astrophysics
Cambridge, MA 02138 USA\and European Space Astronomy Centre, Urb.
Villafranca del Castillo, P.O. Box 50727, Madrid 28080 Spain\and
SOFIA Science Center, Universities Space Research Association, NASA
Ames Research Center, Moffett Field, CA 94035-0001 USA\and Observatorio
Astron\'omico Nacional (IGN), Alfonso XII N$^{\circ}$3, E--28014 Madrid,
Spain\and Observatorio Astron\'omico Nacional (IGN), Ap 112, E--28803
Alcal\'a de Henares, Spain\and Sterrenkundig Instituut Anton Pannekoek,
University of Amsterdam, Science Park 904, NL-1098 Amsterdam, The
Netherlands \and Department of Astrophysics/IMAPP, Radboud University
Nijmegen, Nijmegen, The Netherlands\and Onsala Space Observatory,
Dept. of Radio and Space Science, Chalmers University of Technology,
SE--43992 Onsala, Sweden\and The Netherlands and Astronomical Institute,
Utrecht University, Princetonplein 5, 3584 CC Utrecht, The Netherlands\and
Department of Astronomy, AlbaNova University Center, Stockholm University,
SE--10691 Stockholm, Sweden\and Joint ALMA Observatory, El Golf 40,
Las Condes, Santiago, Chile\and N. Copernicus Astronomical Center,
Rabia\'nska 8, 87-100 Toru\'n, Poland }

\date{Received May 26, 2010; accepted }

\abstract{Water vapour maser emission from evolved oxygen-rich stars remains
poorly understood. Additional observations, including polarisation
studies and simultaneous observation of different maser transitions
may ultimately lead to greater insight.}{We have aimed to elucidate
the nature and structure of the VY CMa water vapour masers in part
by observationally testing a theoretical prediction of the relative
strengths of the 620.701 GHz and the 22.235 GHz maser components of
ortho H$_{2}$O.}{In its high-resolution mode (HRS) the Herschel
Heterodyne Instrument for the Infrared (HIFI) offers a frequency resolution
of 0.125 MHz, corresponding to a line-of-sight velocity of 0.06 km
s$^{-1}$, which we employed to obtain the strength and linear polarisation
of maser spikes in the spectrum of VY CMa at 620.701 GHz. Simultaneous
ground based observations of the 22.235 GHz maser with the Max-Planck-Institut
f\"ur Radioastronomie 100-meter telescope at Effelsberg, provided a
ratio of 620.701 GHz to 22.235 GHz emission.}{We report the first
astronomical detection to date of H$_{2}$O maser emission at 620.701
GHz. In VY CMa both the 620.701 and the 22.235 GHz polarisation are
weak. At 620.701 GHz the maser peaks are superposed on what
appears to be a broad emission component, jointly ejected from the star. We observed the 620.701 GHz emission at two epochs 21 days apart, both to measure the potential
direction of linearly polarised maser components and to obtain a measure
of the longevity of these components. Although we do not detect
significant polarisation levels in the core of the line, they rise up
to approximately 6\% in its wings.}{}


\maketitle
\titlerunning{VY CMa Submillimeter Maser Polarisation}

\section{Introduction}

VY Canis Majoris is a highly luminous, strongly obscured, variable
supergiant with a high infrared excess. Only $\sim1\,\%$ of the total luminosity is observed at optical wavelengths. The star's distance has been measured to be $D=1.1$ kpc \citep{Choi2008} implying a luminosity $L=3\times10^{5}L_{\odot}$ \citep{Menten2008}.  Three thermally emitted mid-infrared water vapour emission lines seen in spectra obtained with the Short Wavelength Spectrometer (SWS) on the Infrared Space Observatory (ISO) indicate a mean radial velocity
of order $20\pm2$~km~s$^{-1}$ and a $25$~km~s$^{-1}$ H$_{2}$O
outflow velocity \citep{Neufeld1999}, significantly lower than the
32\,km\,s$^{-1}$ velocities that \citet{Reid1978} reported for
1612 MHz OH maser outflow.

The effective temperature of the central star,  $T_{*}=2,800$\,K
\citep{Monnier1999},  combined with distance and inferred luminosity fixes the star's radius at $R_{*}\sim10$\,AU.  Working at 11\,$\mu$m, \citet{Danchi1994} reported a variable photospheric radius ranging from 9.5 to 11 mas, where 10 mas corresponds to a stellar radius of 10 AU.

VY CMa is a strong source of $6_{16}-5_{23}$ 22.235 GHz water vapour maser emission. Recent observations at the Atacama Pathfinder Experiment (APEX) telescope have revealed H$_2$O maser emission also at eight submillimetre frequencies, ranging from 321 to 475 GHz \citep{Menten2008}. Although a theoretical model of \citet{NM1991} predicted that the 22.235 GHz transition should be far more luminous than these submillimetre transitions, \citet{Menten2008} find them to be comparable in flux density, or at most a factor of
$\sim6$ lower.  This suggests a need for further observations that might explain the discrepancies.

One of strongest submillimetre masers predicted by \citet{NM1991}
is due to the $5_{32}-4_{41}$ 620.701 GHz transition of ortho water,
whose photon luminosity was expected to be roughly 16\% that of the
22.235 GHz maser. Given the multiple masers observed at 22.235 GHz
we expected the 620.701 GHz maser of VY CMa to exhibit similar multiple
peaks at photon densities $\sim16\%$ of those exhibited at 22.235
GHz.

\section{HIFI Observations at 620.701 GHz}

The Herschel Heterodyne Instrument for the Far Infrared (HIFI) covers
seven frequency bands, ranging from 488.1 to 1901.8 GHz \citep{de Graauw2010}. The 620.701
GHz radiation of the $5_{32}-4_{41}$ maser of ortho H$_{2}$O is
observed in HIFI's Band 1B, which nominally covers the range from
562.6 to 628.4 GHz. Two channels cover each of the frequency bands,
one sensitive to linearly polarised radiation roughly parallel to
the spacecraft horizontal (H) direction, the other roughly parallel
to the vertical direction (V). For Band 1B the H direction of polarisation
is at an angle of 82.5$^{\circ}$ relative to the spacecraft V axis,
while the V direction of polarisation is at an angle of -7.5$^{\circ}$
to that axis. Our data were obtained in the HIFI high-resolution spectroscopy
(HRS) mode with spectral resolution 0.125 MHz, or line-of-sight velocity resolution 0.06 km s$^{-1}$.

The H and V beams on HIFI are not fully coincident. In Band 1B they
are separated by $\sim19\%$ of the $\sim34.4\arcsec$ full-width-half-power beam diameter; the offset between the two beams is $\sim6.6\arcsec$. VY CMa is a spatially unresolved source at these frequencies \citep{Decin2006}.  In our observations the star was positioned half-way between beam centres, i.e., displaced $\sim3.3\arcsec$ relative to the centre of each beam, although a random $\sim1.8\arcsec$ pointing error can slightly increase alignment uncertainty.

In order to determine the orientation on the sky of any observed linear
polarisation, a source has to be viewed at least at two different
rotation angles relative to the telescope.  The Herschel Space
Observatory \citep{Pilbratt2010},
however, cannot be rotated without producing undesirable thermal drifts.
A rotation is best achieved by observing a target at two epochs separated
by a number of weeks. The further the target lies above the ecliptic
plane, the faster is the rotation produced. In our observations, an
interval of three weeks between two sightings of VY CMa resulted in
a rotation of the telescope of $\sim16^{\circ}$ about the line of
sight to the star.

We first observed the star on March 21, 2010 in two contiguous segments,
lasting 10,139 seconds apiece, for a total observing time of 5.633
hours. The first segment began at 08:48:38.0 UT; the second was started
at 11:45:00 UT, 7 minutes and 23 seconds after the first had terminated.
Our second epoch of observations started at 17:02:33 UT on April 11,
2010 and again ran for 10,139 seconds or 2.816 hours.   The system temperature 
for these observations ranged between approximately 80 K and 100 K
with respective RMS noise levels for the first and second sets of
observations of 6.1 and 5.3 Jy for the H channel, and 4.3 and 4.5 Jy
for the V channel at the resolution of 0.125 MHz.

\section{Polarisation Analysis of the 620.701 GHz Maser Components}

Linear polarisation measurements are often optimised to determine
the Stokes $Q$ and $U$ parameters independently from each other
\citep{Li2008,Hezareh2010}. If we define four intensity measurements
$I_{\varphi}$ within the equatorial system with $\varphi$ set to
$0^{\circ}$ when pointing north and increasing eastwards, then \begin{eqnarray}
I & = & \left(I_{0}+I_{45}+I_{90}+I_{135}\right)/2\label{eq:Stokes_I}\\
Q & = & I_{0}-I_{90}\label{eq:Stokes_Q}\\
U & = & I_{45}-I_{135}.\label{eq:Stokes_U}\end{eqnarray}

\noindent Although the state of linear polarisation is completely
defined with these equations, it is also commonly expressed with the
polarisation fraction $p=\sqrt{Q^{2}+U^{2}}/I$ and angle $\theta=0.5\arctan\left(U/Q\right)$.
Alternatively, we can also write \begin{eqnarray}
Q & = & pI\cos\left(2\theta\right)\label{eq:Qcos}\\
U & = & pI\sin\left(2\theta\right).\label{eq:Qsin}\end{eqnarray}

With HIFI we do not generally have access to four independent measurements
precisely sampled with a spacing of $45^{\circ}$, as in Equations
(\ref{eq:Stokes_I})-(\ref{eq:Stokes_U}). Instead, we may have access
to $N$ irregularly spaced observing epochs with the HIFI vertical
polarisation axis (V) oriented at an angle $\varphi_{i}$ relative
to north ($i=1,$2,$\ldots,N$) and the horizontal polarisation axis
(H) at $\varphi_{i}+90^{\circ}$, each observation yielding a pair
of intensities $I_{V_{i}}$ and $I_{H_{i}}$.

These $2N$ intensities yield the polarisation if we first define
\begin{eqnarray}
d_{i} & \equiv & I_{V_{i}}-I_{H_{i}}\nonumber \\
 & = & Q\cos\left(2\varphi_{i}\right)+U\sin\left(2\varphi_{i}\right).\label{eq:d_i}\end{eqnarray}

\noindent Introducing a vector $\mathbf{d}$ with elements $d_{i}/\sigma_{i}$,
where $d_{i}$ is the intensity difference of Equation (\ref{eq:d_i})
and $\sigma_{i}$ its uncertainty (i.e., the corresponding noise intensity),
we can write the matrix equation \begin{equation}
\mathbf{d}=\mathbf{A}\cdot\mathbf{s},\label{eq:d_vector}\end{equation}

\noindent where the matrix $\mathbf{A}$ and vector $\mathbf{s}$
are given by \begin{eqnarray}
\mathbf{A} & = & \left(\begin{array}{cc}
\cos\left(2\varphi_{1}\right)/\sigma_{1} & \sin\left(2\varphi_{1}\right)/\sigma_{1}\\
\vdots & \vdots\\
\cos\left(2\varphi_{N}\right)/\sigma_{N} & \sin\left(2\varphi_{N}\right)/\sigma_{N}\end{array}\right)\label{eq:A_matrix}\\
\mathbf{s} & = & \left(\begin{array}{c}
Q\\
U\end{array}\right).\label{eq:s_vector}\end{eqnarray}

The vector $\mathbf{s}$, and therefore the Stokes parameters, are
easily obtained by inverting Equation (\ref{eq:d_vector}) using the
pseudo-inverse matrix of $\mathbf{A}$. We then find \begin{equation}
\mathbf{s}=\left[\left(\mathbf{A}^{T}\mathbf{A}\right)^{-1}\mathbf{A}^{T}\right]\cdot\mathbf{d}.\label{eq:s_solution}\end{equation}

The elements of the covariance matrix $\mathbf{C}=\left(\mathbf{A}^{T}\mathbf{A}\right)^{-1}$
can be shown to yield the corresponding uncertainties on the estimated
Stokes parameters with \begin{equation}
\mathbf{C}=\left(\begin{array}{cc}
\sigma_{QQ}^{2} & \sigma_{QU}^{2}\\
\sigma_{QU}^{2} & \sigma_{UU}^{2}\end{array}\right),\label{eq:C_uncertainties}\end{equation}

\noindent where $\sigma_{ij}^{2}=\left\langle s_{i}s_{j}\right\rangle -\left\langle s_{i}\right\rangle \left\langle s_{j}\right\rangle $
with $i,j=Q$ or $U$ and $s_{i}$ the appropriate component of the
vector $\mathbf{s}$. It is also possible to determine the uncertainties
$\sigma_{p}$ and $\sigma_{\theta}$ for the polarisation fraction
and angle, respectively, with \begin{eqnarray}
\sigma_{p}^{2} & \simeq & \frac{1}{p^{2}I^{4}}\left(Q^{2}\sigma_{QQ}^{2}+U^{2}\sigma_{UU}^{2}+2QU\sigma_{QU}^{2}\right)\label{eq:sigma_p}\\
\sigma_{\theta}^{2} & \simeq & \frac{1}{4p^{4}I^{4}}\left(Q^{2}\sigma_{UU}^{2}+U^{2}\sigma_{QQ}^{2}-2QU\sigma_{QU}^{2}\right).\label{eq:sigma_theta}\end{eqnarray}

\begin{figure}
\includegraphics[scale=0.5]{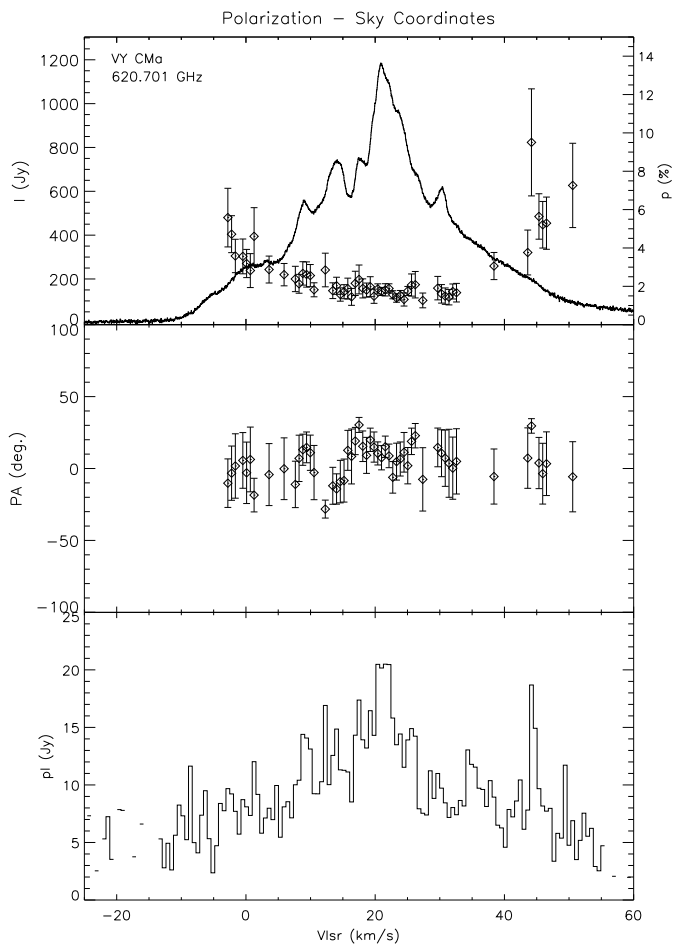}
\caption{\label{fig:spectra}Polarisation spectrum of the 620.701 GHz $\mathrm{H}_{2}\mathrm{O}$
$5_{32}-4_{41}$ ortho-transition for VY-CMa. Shown are, from top
to bottom, the Stokes $I$ and polarisation fraction (symbols and right
axis), the polarization
angle, and the (unsigned) polarised flux. The polarisation fraction and flux were corrected for positive
bias but not for instrumental polarisation. A simple linear baseline was removed from each $V$ and $H$
channel observation and all polarisation vectors shown satisfy $p\geq3\sigma_{p}$
and $I\geq10\sigma_{I}$ at the corresponding velocities, where $\sigma_{I}$
is the uncertainty (or noise level) on the Stokes $I$ spectrum (i.e.,
the average of the $I_{V}+I_{H}$ spectra measured at the two epochs).
The polarisation fraction and angle have their spectral resolution reduced by a factor of 20 compared to Stokes $I$, which is at the maximum resolution.}

\end{figure}

We have applied this polarisation analysis to our observations of
the 620.701 GHz $\mathrm{H}_{2}\mathrm{O}$ $5_{32}-4_{41}$ ortho-transition.
Our results shown in Figure \ref{fig:spectra} were obtained from
the two aforementioned observations made at epochs of corresponding
position angles of $261.27^{\circ}$ and $277.46^{\circ}$. 

Although there are no obvious strong polarisation signals from the
maser emission peaks, we clearly detect polarisation levels ranging
from $p\simeq1.5\%$ to $p\simeq 6\%$ in regions of significant line
intensity (i.e., from approximately -5 to 45 km~s$^{-1}$). Furthermore,
the observed anti-correlation of the polarisation fraction with the
Stokes $I$ intensity is similar to previous ground-based polarisation
observations \citep{Girart2004,Hezareh2010} aimed at the detection
of the Goldreich-Kylafis effect in non-masing molecular lines \citep{GK1981,Cortes2005}, which appears to have first been detected in evolved stars \citep{Glenn1997}. We will discuss the relevance of the Goldreich-Kylafis effect for our observations in section 7 below.

\section{Results}

Several instrumental capabilities of HIFI and the observations obtained
with them may be noted:

(i) The two orthogonally polarised HIFI receivers are well matched
and extremely stable. But observations of
extended sources need to be conducted with caution.  A slight
misalignment of H and V receivers can lead to a ``false polarisation"
that reverses polarity at half-year intervals  (see Appendix \ref{sec:1557}).

(ii) The misalignment of the HIFI receivers does not appear to affect
observations of unresolved sources. Our observations realised at
two observing epochs indicate that instrumental polarisation, which
could be in part due to errors in the relative calibration between
the two receiver chains of Band 1B, cannot exceed a measure of order $1-2\%$ (see section 7 below).

(iii) The polarisation of VY CMa is not significant
near the peak of the 620.701 GHz $\mathrm{H}_{2}\mathrm{O}$ $5_{32}-4_{41}$
line, but rises up to $\sim 6\%$ in the wings of the spectrum in
a manner consistent with polarisation due to the Goldreich-Kylafis
effect discussed in greater detail in section 7 below.

(iv) The stability of the 620.701 GHz masers is remarkable. The variation
over a three week period is $\lesssim1\%$ (see Figure 2).

\begin{figure}
\resizebox{\hsize}{!}{\rotatebox{270}{\includegraphics{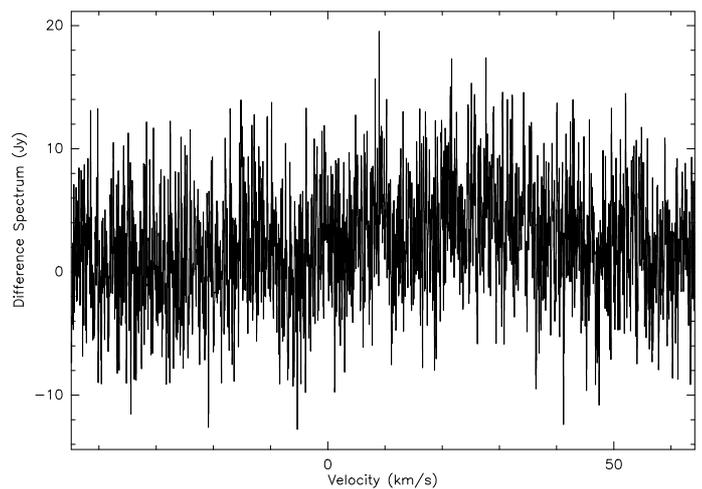}}} 
\caption{\label{fig:diff_spectrum}Residual spectrum resulting from the
  subtraction of the 620.701 GHz Stokes $I$ spectra taken 3 weeks apart. The residual signal has an amplitude $\lesssim1\%$, underlining the accuracy of the relative calibrations as well as the remarkable stability of the masers. }

\end{figure}

\begin{figure}
\resizebox{\hsize}{!}{\includegraphics{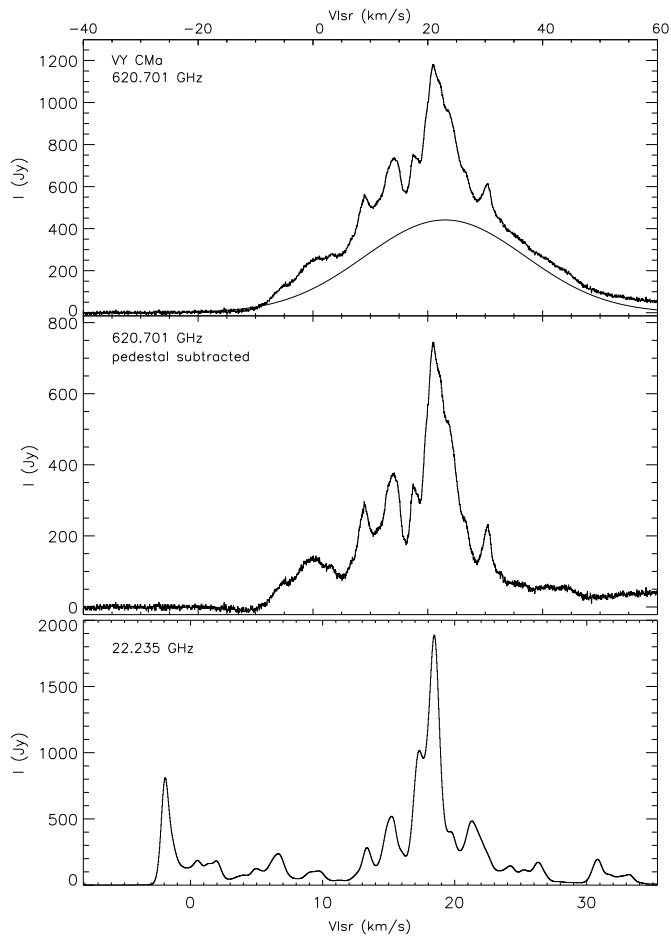}} 
\caption{\label{fig:22vs621}Comparison of the VY CMa 620.701 and 22.235 GHz
masers, both observed on April 11, 2010. The 620.701 GHz {\it(top)} and the 22.235 GHz {\it(bottom)} spectra have different velocity scales, and their peaks fall at slightly different $v_{\rm{lsr}}$ velocities that have been shifted to coincide.   Superposed on the 620.701 GHz spectrum is a Gaussian pedestal subtracted in the spectrum just below to display the similarities of the 620.701 and  22.235 GHz maser spectra.  Not knowing the source or true shape of the pedestal, we kept its spectrum simple.}

\end{figure}

(v) As Figure \ref{fig:22vs621} shows, the spectral profile of the
620.701 GHz and 22.235 GHz masers appears remarkably similar, except that the relative
expansion velocities between peaks, along the line of sight, is a
factor of 2.3 greater at 620.701 than for contemporaneous
  observations at 22.235 GHz (see Appendix
\ref{sec:22GHz}).

(vi) Assuming the 620.701 GHz masers to sit atop a broad pedestal with
flux density $\sim400$ Jy, we find the main 620.701 GHz maser peak
flux density to be $\sim800$ Jy, compared to $\sim1900$ Jy at 22.235
GHz.  The velocity spread of the expanding 620.701 GHz maser peaks is $\sim2.3$ times wider than at 22.235 GHz, implying a 620.701 GHz photon luminosity roughly 2.4 times lower than at 22.235 GHz --- comparable in luminosity to the submillimetre masers observed by \citet{Menten2008}, and about 2.5 times more luminous than \citet{NM1991} predicted.

\section{Discussion}

A spherically symmetric model for the outflows of VY CMa developed by \citet{Decin2006}
  envisions dust formation in the cooling outflow from the star at
  distances $\sim 10 R_*$.  Radiation pressure then accelerates the
  dust, and with it also the ambient gas, from velocities $\lesssim 5$ km
  s$^{-1}$ to $\sim 25$ km s$^{-1}$ at a distance of $20 R_*$. The
  kinetic temperature at $20 R_*$  is of the order of $\sim 1000$ K,
  sufficient to excite both 620.701 and 22.235 GHz masers.  We propose
  that the masing outflows we observe propagate along several mutually
  common directions from the star, and that the faster expanding
  620.701 GHz masers lie at greater distances from the star than the
  central cluster of masers peaking at 22.235 GHz. We picture this outflow consisting of gas ejected obliquely to the line of sight to
  VY CMa but fanning out over a considerable angular width to account for the observed range of outflow velocities. Common ejection of both the denser masing regions and the less-dense ambient medium responsible for the pedestal emission can then account for the observed velocity ranges of both these outflow components.  The region from which the 22.235 GHz radiation originates may be at the inner radius of the flow; for, \citet{NM1991} find that, for comparable gas densities and H$_2$O abundances, maser emission at 22.235 GHz should be maximized at line-of-sight velocity gradients $\sim 30$ times lower than those maximizing 620.701 GHz emission.

A comparison of our 620.701 GHz spectrum to the spectra of the star's submillimeter masers observed by \citet{Menten2008} shows that their spectra, observed in June 2006, four years earlier than ours, generally exhibited only two peaks separated by $\lesssim 6$ km s$^{-1}$, whereas our spectrum at 620.701 GHz clearly exhibits four peaks ranging over separations of order $\sim 25$ km s$^{-1}$.  However, their 22.235 GHz spectrum also looks quite different from ours.  This is not surprising since \citet{Esimbek2001} found that, between August 1993 and August 1999, the shape of the 22.235 GHz spectrum significantly changed and peak emission shifted from  $v_{\rm{lsr}}\sim 17$ to 34.2 km s$^{-1}$.  

 Nevertheless, a feature common to all the submillimetre spectra of \citet{Menten2008} is a broad pedestal that diminishes in strength with rising upper-excitation temperature $E_u$ of the observed transitions.  For $E_u\sim 725$ K it remains strong with wings extending over a full width $\gtrsim 40$ km s$^{-1}$, whereas at $E_u$ = 1861 K, the pedestal has entirely disappeared. The radiation emitted in the pedestal thus appears to be due to collisional excitation by gas flowing out from the star at speeds $\gtrsim 20$ km s$^{-1}$ close to the terminal velocity in the model of \citet{Decin2006}. The pedestal we observe for the transition at 620.701 GHz, whose $E_u = 732$ K, has the same broad wings extending roughly symmetrically to either side of the maser outflow peaks.

Low $E_u$ levels thus excited can then radiate in three ways,
spontaneously, through stimulated emission, or induced by
collisions. As \citet{NM1991} point out, stimulated emission leads to
masing only when the ratio of water vapor density to the velocity
gradient along the line of sight is sufficiently high.  We interpret
the  maser peaks in our 620.701 GHz spectrum as emanating from denser
clumps in the VY CMa outflow within which the velocity gradient is
low, and the 620.701 GHz pedestal as emission from an ambient medium
characterized by lower density and/or higher velocity gradients, and
thus unable to radiate significantly through stimulated emission to
sustain maser amplification.  This ambient medium, and its
corresponding radiation, will bear the imprint of the
Goldreich-Kylafis (GK) effect when also subjected to anisotropic
radiation or optical depths, and permeated by a weak magnetic field.

Although the low polarization levels we detect are expected for water
masers \citep{Watson2009}, the rise in polarisation seen in the wings of
the spectrum in the top panel of Figure \ref{fig:spectra} deserves careful
consideration. If we assume that the masers simply maser-amplify
the seed radiation emanating from the ambient gas responsible for the
pedestal component, then we may have a qualitative explanation for the shape
of the polarisation spectrum. That is, the polarization fraction
should be approximately the same at the position of, and between,
maser peaks, as observed in Figure \ref{fig:spectra}, and
should rise in the wings of the line where the optical depth of the
pedestal component is lowest, in a manner consistent with the GK
effect \citep{Cortes2005}. Although maser-amplification of GK polarisation could lead to
different levels of polarisation at the maser peaks, other factors
(e.g., magnetic field orientation) could also suppress this effect. We
also note that in cases where the pedestal component was highly
saturated, the level of polarisation measured in its core would
provide a measure of the amount of instrumental polarisation
\citep{Hezareh2010}. Although this scenario probably does not
perfectly fit our observations, the polarisation fraction measured in the vicinity of the line core suggests that the instrumental polarisation is approximately 1 - 2\%.

\section{Conclusions}

We report the first astronomical observations of the 620.701 GHz $5_{32}-4_{41}$ submillimetre maser of ortho H$_{2}$O.  The maser peaks in VY CMa show weak linear polarisation at levels consistent with the Goldreich-Kylafis effect across the spectral line profile.

\begin{acknowledgements}
HIFI has been designed and built by a consortium of institutes and
university departments from across Europe, Canada and the United States
under the leadership of SRON Netherlands Institute for Space Research,
Groningen, The Netherlands and with major contributions from Germany,
France and the US. Consortium members are: Canada: CSA, U.Waterloo;
France: CESR, LAB, LERMA, IRAM; Germany: KOSMA, MPIfR, MPS; Ireland,
NUI Maynooth; Italy: ASI, IFSI-INAF, Osservatorio Astrofisico di Arcetri-
INAF; Netherlands: SRON, TUD; Poland: CAMK, CBK; Spain: Observatorio
Astronmico Nacional (IGN), Centro de Astrobiolog\'ia (CSIC-INTA). Sweden:
Chalmers University of Technology - MC2, RSS \& GARD; Onsala Space
Observatory; Swedish National Space Board, Stockholm University -
Stockholm Observatory; Switzerland: ETH Zurich, FHNW; USA: Caltech,
JPL, NHSC. Support for this work was provided by NASA through an award
issued by JPL/Caltech. We thank the HIFISTARS consortium for permission
to use their VY CMa 556.9 GHz data for calibration purposes.  We would like to acknowledge that Tom Phillips first pointed out to us that HIFI might yield useful polarisation observations.  And, finally, we extend thanks to the anonymous referee for perceptive insights and suggestions that greatly improved this paper.
\end{acknowledgements}

\appendix

\section{\label{sec:1557}A Cautionary Note on Observations of Extended Sources}

As part of the Performance Verification of HIFI in space, spectroscopic
observations of the $^{12}$C$^{16}$O $J=5\rightarrow4$ 576.3 GHz transition
were obtained at the position of the chemically-active bow shock B1
driven by the LDN 1157 Class 0 protostar \citep{Bachiller2001}.  Initial  observations in HIFI's Band 1B were made on August 1, 2009; a second set was obtained 186 days later on February 4, 2010. The spectral shape of the source remained essentially unchanged but initially the signal from the high-frequency edge of the line appeared stronger in the V direction, when the H and V peaks were matched, whereas half a year later it appeared stronger by the same ratio in the H direction.

This ``false polarisation" effect was due to the small misalignment of
the H and V beams described earlier, which imaged slightly
different portions of a scene onto the two receivers \citep{Attard2008}.
Unless otherwise specified, observations of a targeted source are
centred on a position halfway between the two beam centres. Because
the viewing direction of Herschel at all times is constrained by the
need to keep the plane of the sun-shield roughly perpendicular to
the radius vector to the Sun, sources close to the ecliptic plane
can only be viewed at half year intervals. Over this interval the
telescope aperture rotates 180$^{\circ}$ about its viewing direction
on the sky. Thus, portions of LDN 1157 B1 initially viewed
by the V channel were viewed, half a year later, by the H channel
and vice versa. It is clear that this effect was not due to linear
polarisation because a 180$^{\circ}$ rotation of the telescope leaves
the polarisation direction unchanged.

For VY CMa, where the entire source was relatively well centred on each beam,
this effect was negligible.  But for linear polarisation observations of an extended source the best strategy may be to make a small map with HIFI over the area of interest, where the spectra obtained in the H and V channels at each position of the source, can be individually and directly compared. 

\section{\label{sec:22GHz}Ground-based Observations at 22.235 GHz}

Overlapping with the second epoch of 620.701 GHz observations, we
observed the VY CMa 22.235 GHz maser with the Effelsberg 100-meter
telescope of the Max-Planck-Institut f\"ur Radioastronomie. Those
observations were begun on April 11, 2010 at 16:40 UT and lasted until
17:20 UT  and are shown
  in Figure \ref{fig:22vs621}. Data in two linear polarisations were obtained with the two channels of the K-band or 1.3 cm receiver at the primary focus. The frequency resolution was 6.104 kHz, corresponding to a velocity resolution of 0.082 km s$^{-1}$. Although these observations were realised at only one epoch and are, therefore, insufficient for  a full characterisation of the linear polarisation state, a search by \citet{Vlemmings2002} found  that the linear polarisation of the VY CMa 22.235 GHz  masers is well below 1\%.

\end{document}